# Orbital driven two-dome superconducting phases in multiorbital superconductors

Jing Liu[1,2,*], Qing-Wei Wang[3,4,*], and Liang-Jian Zou [1,2,#]

[1]*Key Laboratory of Materials Physics, Institute of Solid State Physics, Chinese Academy of Sciences, P. O. Box 1129, Hefei 230031, China*

[2] *Science Island Branch of Graduate School, University of Science and Technology of China, Hefei 230026, China*

[3] *School of Mathematics, Physics & Information Science, ZheJiang Ocean University, Zhoushan, Zhejiang 316022, China*

[4] *Key Laboratory of Oceanographic Big Data Mining & Application of Zhejiang Province, ZheJiang Ocean University, Zhoushan, Zhejiang 316022, China*

**Abstract**

We theoretically study the superconductivity in multiorbital superconductors based on a three-orbital tight-banding model. With appropriate values of the nearest-neighbour exchange $J_1^{\alpha\beta}$ and the next-nearest-neighbour exchange $J_2^{\alpha\beta}$, we find a two-dome structure in the $T_c - n$ phase diagram: one dome in the doping range $n < 3.9$ where the superconducting (SC) state is mainly $s_{x^2y^2}$ component contributed by inter-orbital pairing, the other dome in the doping range $3.9 < n < 4.46$ where the SC state is mainly $s_{x^2y^2} + s_{x^2+y^2}$ components contributed by intra-orbital pairing. We find that the competition between different orbital pairing leads to two-dome SC phase diagrams in multiorbital superconductors, and different matrix elements of $J_1$ and $J_2$ considerably affect the boundary of two SC domes.

* These authors equally contribute this manuscript.
# Corresponding author:zou@theory.issp.ac.cn

# 1. Introduction

The discovery and development of iron-based superconductors (IBSC) [1-4] provide new opportunities and bring great challenges in the quest for unconventional superconducting mechanisms in these high-$T_c$ correlated electron systems. Though the IBSC display great resemblance with cuprate superconductors in their high-temperature superconductivity and magnetism, there exist important differences between these two families of superconductors. First of all, the parent compounds of the IBSC are usually bad metals with the magnetic structure of stripe spin-density wave [5], while the undoped cuprates are half-filled antiferromagnetic insulators. Second, the multiorbital or multi-band nature of the IBSC [6-19] is crucially different from the effective single-band nature of the cuprate superconductors [20]. Third, the IBSC are intermediately correlated, with the next-nearest-neighbour (NNN) hoppings and exchanges between iron sites always dominating over the nearest-neighbour (NN) ones [10,16-19]; while, the cuprates are strongly short-range correlated, with the NNN hoppings and exchanges much smaller than the NN ones. It leads to a distinct consequence for the SC gap symmetry: the IBSC tend to $s^{+-}$ or $s^{++}$ wave symmetry [9-19], while the cuprates tend to $d$ wave to reduce the Coulomb repulsion.

The coexistence and competition between different orbitals and between the NN and NNN interactions contribute many exotic properties. For example, the orbital selective Mott physics was suggested in some IBSC [21,22]. Especially, unique two-dome superconductivity was found, $i.e.$, after the $T_c$ reaches the maximum and then decreases to vanish in the first superconducting (SC) dome, a second SC dome appears, close to or separated from the first one. Such two-dome SC phase diagrams have been widely observed in doped $LaFeAsO_{1-x}H_x$ [23], $SmFeAs_{1-y}P_yO_{1-x}H_x$ [24], $LaFeAsO_{1-x}F_x$ [25], $PrFeAsO_{0.6}F_y$ [26], $LaFeP_{1-x}As_xO_{1-y}F_y$ [27], and $K_xFeSe$ ultrathin films [28] with increasing doping, as well as in pressured $FeS$ [29], $FeSe_{1-x}$ [30,31], $RFe_2As_2$ ( $R = K, Cs, Rb$ ) [32-34], $R_yFe_{2-x}Se_2$ ( $R = K, Tl, Rb$ ) [35], $(Li_{1-x}Fe_x)OHFe_{1-y}Se$ [36], $FeSe_{0.89}S_{0.11}$ [37] and $(NH_3)_yCs_xFeSe$ [38]. In fact, the two-dome SC phase diagram has also been found in other unconventional superconductors, including the heavy fermion [39-43],copper-oxides [44-49], organics [50,51] and oxybismuthides [52], some of which have been summed in Ref. [53].The two-dome SC phase may or may not accompany with bipartite magnetic phases [54-56]. To study the two-dome SC phase may provide a hint for pursuing higher $T_c$ SC and disclosing the SC pairing-mechanism of unconventional SC.

According to Hosono $et\ al.$ [23] two-dome SC phase can be theoretically interpreted using the electronic structure as follows: the first dome is ascribed to the nesting between the hole and electron pockets, while the peak of density of state (DOS) of electrons give rise to the second dome. Suzuki $et\ al.$ [57] suggested that this characteristics is due to the competition between the NN and NNN hoppings across iron sites. Whereas, Liu $et\ al.$ [9,56] associates these characteristics with the multiorbital nature of IBSC. Though the SC properties of the IBSC could be affected by doping or pressure in several ways, such as (i) the crystal and valence bond structure, (ii) the electron-hole number density in the iron arsenide (FeAs) layers, (iii) the hopping integrals and spin exchange interactions between iron sites, and so on. It remains a controversial topic that which factor plays a decisive role in IBSC. Further investigation is important for understanding the essence of IBSC.

In this paper, starting from a three-orbital tight-binding (TB) model to describe the unconventional SC in multiorbital compounds, we study the SC order parameters as the functions of electron occupation $n$. We find that the SC order along with the critical temperature $T_c$ as the functions of the electron occupation $n$ displays a clear two-dome structure, and the analysis of the electronic state near Fermi surface (FS) shows that the variation of orbital states contributes this unusual phase diagram. The rest of this paper is organized as follows: we first outline the model Hamiltonian and the self-consistent mean-field equations for the SC order parameters in Sec.2. The numerical results including the phase diagram, the electronic structure near the FS are presented in Sec.3. Finally we draw our conclusions in Sec.4.

## 2. Model Hamiltonian and formula

We start from a three-orbital TB model with the NN and NNN exchange interactions and Hund's rule coupling [9,10], which reads

$$H_0 = \sum_{\mathbf{k}\alpha\beta\sigma} T^{\alpha\beta}(\mathbf{k}) d^+_{\mathbf{k}\alpha\sigma} d_{\mathbf{k}\beta\sigma}$$

$$+ \sum_{\mathbf{r}\delta\alpha\beta} J_1^{\alpha\beta} \left[ \mathbf{S}(\mathbf{r},\alpha) \cdot \mathbf{S}(\mathbf{r}+\delta,\beta) - \frac{1}{4} n(\mathbf{r},\alpha) n(\mathbf{r}+\delta,\beta) \right]$$

$$+ \sum_{\mathbf{r}\tilde{\delta}\alpha\beta} J_2^{\alpha\beta} \left[ \mathbf{S}(\mathbf{r},\alpha) \cdot \mathbf{S}(\mathbf{r}+\tilde{\delta},\beta) - \frac{1}{4} n(\mathbf{r},\alpha) n(\mathbf{r}+\tilde{\delta},\beta) \right]$$

$$+ \sum_{\mathbf{r},\alpha\neq\beta} J_H \mathbf{S}(\mathbf{r},\alpha) \cdot \mathbf{S}(\mathbf{r},\beta), \tag{1}$$

where the operator $d^+_{\mathbf{k}\alpha\sigma}$ $(d_{\mathbf{k}\alpha\sigma})$ creates (annihilates) an electron with momenta $\mathbf{k}$, orbital index $\alpha = xz/yz/xy$ and spin $\sigma$, $\delta$ or $\tilde{\delta}$ is the position vector of the NN or NNN site. The matrix elements of the hopping matrix $T^{\alpha\beta}(\mathbf{k})$ are [10]

$$T^{11} = 2t_2\cos k_x + 2t_1\cos k_y + 4t_3\cos k_x\cos k_y,$$

$$T^{22} = 2t_1\cos k_x + 2t_2\cos k_y + 4t_3\cos k_x\cos k_y,$$

$$T^{33} = 2t_5(\cos k_x + \cos k_y) + 4t_6\cos k_x\cos k_y + \delta_{xy},$$

$$T^{12} = T^{21} = 4t_4\sin k_x\sin k_y,$$

$$T^{13} = T^{31} = 2it_7\sin k_x + 4it_8\sin k_x\cos k_y,$$

$$T^{23} = T^{32} = 2it_7\sin k_y + 4it_8\sin k_y\cos k_x, \tag{2}$$

where the NN and NNN hopping integrals $t_1 = 0.02$, $t_2 = 0.06$, $t_3 = 0.03$, $t_4 = -0.01$, $t_5 = 0.2$, $t_6 = 0.3$, $t_7 = -0.2$, and $t_8 = t_7/2$, taking electron volt as the energy unit. $J_H$ is the Hund's rule coefficient. $J_1^{\alpha\beta}$ ($J_2^{\alpha\beta}$) is the NN (NNN) spin coupling coefficient. The $xz$ and $yz$ orbitals are degenerate and the $xy$ orbital is slightly split from the $xz$ and $yz$ orbitals about $\delta_{xy}$=0.4. This Hamiltonian, which gives the correct Fermi-surface topology of LaOFeAs, represents a typical model of three-orbital compounds on a square lattice, though the details of Fermi surfaces depend on concrete FeAs-based compounds.

The spin fluctuations, possible as well as the orbital fluctuations, give rise to the SC pairing forces in these unconventional SC [58,59]. In the multiorbital SC state, we define the SC order parameters in real space as follows:

$$\Delta^{\alpha\beta}(\mathbf{r}) = -J_{\mathbf{r}}^{\alpha\beta}\left[\phi_{\alpha\beta}(\mathbf{r}) + \phi_{\beta\alpha}(-\mathbf{r})\right], \quad J_{\mathbf{r}}^{\alpha\beta} = \begin{cases} J_1^{\alpha\beta}, \text{when } \mathbf{r} = \delta; \\ J_2^{\alpha\beta}, \text{when } \mathbf{r} = \widetilde{\delta}; \\ 0, \text{ otherwise.} \end{cases} \quad (3)$$

where $\phi_{\alpha\beta}(\mathbf{r}) = \frac{1}{N}\sum_{\mathbf{k}}\langle d_{-\mathbf{k},\alpha\downarrow}d_{\mathbf{k}\beta\uparrow}\rangle e^{-i\mathbf{k}\cdot\mathbf{r}}$, $N$ denotes the total number of Fe sites. Then in $\mathbf{k}$ space,

$$\Delta^{\alpha\beta}(\mathbf{k}) = \sum_{\mathbf{r}}\Delta^{\alpha\beta}(\mathbf{r})e^{-i\mathbf{k}\cdot\mathbf{r}} \approx \sum_{\delta}\Delta^{\alpha\beta}(\delta)e^{-i\mathbf{k}\cdot\delta} + \sum_{\widetilde{\delta}}\Delta^{\alpha\beta}(\widetilde{\delta})e^{-i\mathbf{k}\cdot\widetilde{\delta}}.$$

Meanwhile, we define the homogeneous potential $V_c = V_c^+$ due to the charge density:

$$V_c^{\alpha\alpha} = -4\sum_{\beta}\left(J_1^{\alpha\beta} + J_2^{\alpha\beta}\right)n_{\beta\beta}, \quad (4a)$$

$$V_c^{\alpha\beta} = \frac{3}{2}J_H n_{\beta\alpha}, \ (\alpha \neq \beta), \quad (4b)$$

where $n_{\alpha\beta} = \frac{1}{N}\sum_{\mathbf{k}}\langle d_{\mathbf{k}\alpha\sigma}^+ d_{\mathbf{k}\beta\sigma}\rangle$. Then in the SC phase the Hamiltonian (1) can be decoupled in $\mathbf{k}$ space. After a standard mean-field process [9], we get

$$H_{MFT} = \sum_{\mathbf{k}}\Psi^+(\mathbf{k})A(\mathbf{k})\Psi(\mathbf{k}) + NE_0, \quad (5)$$

where $A(\mathbf{k}) = \begin{bmatrix} T(\mathbf{k}) + V_c - \mu & \Delta(\mathbf{k}) \\ \Delta^+(\mathbf{k}) & -T(\mathbf{k}) - V_c + \mu \end{bmatrix}$,

$$E_0 = \frac{1}{2}\sum_{\delta\alpha\beta}J_1^{\alpha\beta}\left|P^{\alpha\beta}(\delta)\right|^2 + \frac{1}{2}\sum_{\widetilde{\delta}\alpha\beta}J_2^{\alpha\beta}\left|P^{\alpha\beta}(\widetilde{\delta})\right|^2 + 4\sum_{\alpha\beta}\left(J_1^{\alpha\beta} + J_2^{\alpha\beta}\right)n_{\alpha\alpha}n_{\beta\beta}$$

$$+ \text{Tr}V_c - (3-n)\mu + \Delta_{xy},$$

$\Psi^+(\mathbf{k}) = (d_{\mathbf{k}1\uparrow}^+, d_{\mathbf{k}2\uparrow}^+, d_{\mathbf{k}3\uparrow}^+, d_{-\mathbf{k},1\downarrow}, d_{-\mathbf{k},2\downarrow}, d_{-\mathbf{k},3\downarrow})$, $P^{\alpha\beta}(\delta) = -2J_1^{\alpha\beta}\phi_{\alpha\beta}(\delta)$, and $P^{\alpha\beta}(\widetilde{\delta}) = -2J_2^{\alpha\beta}\phi_{\alpha\beta}(\widetilde{\delta})$. Both the order parameters $\Delta(\mathbf{k})$ and $V_c$ are $3 \times 3$ matrices and hence the total matrix $A(\mathbf{k})$ is a $6 \times 6$ matrix. $\mu$ is the chemical potential.

The mean-field Hamiltonian (5) can be diagonalized and one can get the ground-state energy. Minimizing the ground-state energy with respect to the preceding order parameters gives rise to the average of $\phi_{\alpha\beta}$ and $n_{\alpha\beta}$, hence Eq. (3) and (4) are also self-consistent equations for $\Delta^{\alpha\beta}(\mathbf{k})$ and $V_c$, which can be solved numerically, combining with the particle number constraint corresponding to the chemical potential $\mu$. Then we could obtain the dependence of the order parameters $\Delta^{\alpha\beta}(\mathbf{k})$ and $V_c$ on doping, temperature and various physical conditions.

Accordingly, we could analyze the pairing symmetry in the SC state. From the symmetry of the square lattice, the SC order parameter $\Delta(\mathbf{k})$ of the systems can be rewritten as the superposition of various pairing components,

$$\Delta(\mathbf{k}) = (\Delta_x + \Delta_y)(\cos k_x + \cos k_y) + (\Delta_x - \Delta_y)(\cos k_x - \cos k_y)$$
$$+ 2\Delta_{xy}\cos k_x \cos k_y, \qquad (6)$$

where $\Delta_x \equiv \Delta(\mathbf{e}_x)$, $\Delta_y \equiv \Delta(\mathbf{e}_y)$, $\Delta_{xy} \equiv \Delta(\mathbf{e}_x + \mathbf{e}_y)$, $\mathbf{e}_x$ and $\mathbf{e}_y$ are the unit vectors of the crystal lattice. That is, the pairing wave function is a superposition of $s_{x^2y^2} \propto \cos k_x \cos k_y$, $s_{x^2+y^2} \propto (\cos k_x + \cos k_y)$ and $d_{x^2-y^2} \propto (\cos k_x - \cos k_y)$ waves. For different signs on different portions of the FS, the $s_{x^2y^2}$ and $s_{x^2+y^2}$ waves are also referred to as $s^{+-}$ and $s^{++}$ waves, respectively.

In what follows we present the numerical solution of the Hamiltonian (1). The energy band (2) is fitted near $n$=4.0 (a filling of two thirds) [10], so we consider the filling range near $n$=4.0 throughout this paper. The Hund's rule coupling is taken as $J_H = 0.35$ eV according to the estimation in Ref. [13].

## 3. Numerical results

In this section, we first obtain the dependence of the SC order parameters on the electron occupation $n$, as shown in Fig. 1, where the NN and NNN superexchange coefficients $J_1^{\alpha\beta}$ and $J_2^{\alpha\beta}$ are taken as $J_1 = \begin{bmatrix} 0.02 & 0.04 & 0.04 \\ 0.04 & 0.02 & 0.04 \\ 0.04 & 0.04 & 0.04 \end{bmatrix}$ eV and $J_2 = \begin{bmatrix} 0.05 & 0.1 & 0.1 \\ 0.1 & 0.05 & 0.1 \\ 0.1 & 0.1 & 0.1 \end{bmatrix}$ eV, respectively. It shows that the pairing symmetry of the SC order parameters consists mainly of the $s_{x^2y^2}$ and $s_{x^2+y^2}$ waves in or between the $xz$ and $yz$ orbitals: in the filling range of $n < 3.9$, the $s_{x^2y^2}$ wave between the $xz$ and $yz$ orbitals dominates the SC pairing symmetry, as seen the blue triangles in Fig. 1; in the filling range of $3.9 \leq n \leq 4.46$, the $s_{x^2y^2}$ wave intra the $xz$ and $yz$ orbitals dominates the SC pairing symmetry. The SC pairing amplitudes of the $d_{x^2-y^2}$ wave and of the $xy$ orbital channel are not plotted since they are quite small in magnitude.

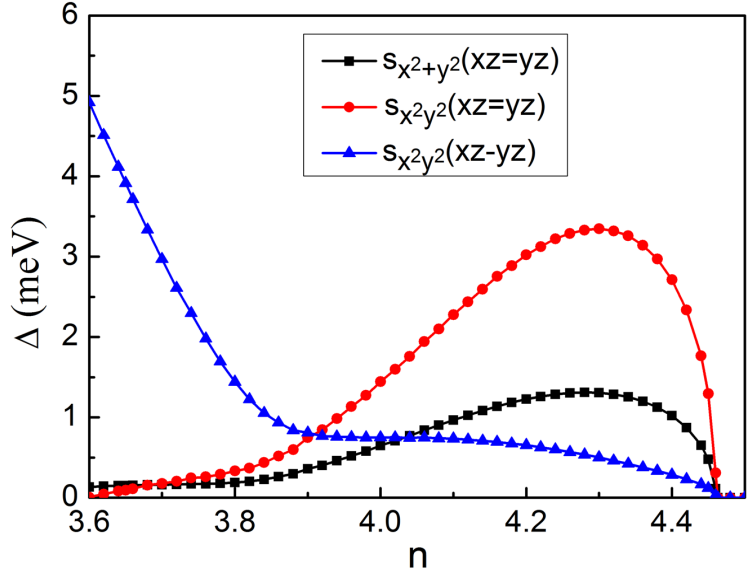

**Fig. 1.** The SC order parameters as functions of electron occupation $n$ at the temperature 5K. The black and red lines represent the intra-orbital pairing which have the same strength in the $xz$ and $yz$ orbitals (= indicates equally). The blue line represents the inter-orbital pairing between the $xz$ and $yz$ orbitals ($xz$-$yz$).

As expected, the two-dome structure of the SC order is obtained in the $n - T_c$ phase diagram, as shown in Fig. 2: with the increase of $n$, $T_c$ first monotonously decreases to a minimum at $n = 3.9$; then it continuously increases to a maximum; with further increasing $n$, $T_c$ quickly decreases to vanish around $n = 4.46$. The maximum of the inter-orbital pairing at $n = 3.6$ corresponds to the maximum $T_c = 70K$; the maximum of the intra-orbital pairing appears at $n = 4.3$ with the maximum $T_c = 54K$. This two-dome structure is widely seen in the experimental phase diagram of H-doped LaFeAsO [54], F-doped LaFeAsO [25] and other multiorbital IBSC [35], suggesting that the orbital variation drives two-dome SC structure.

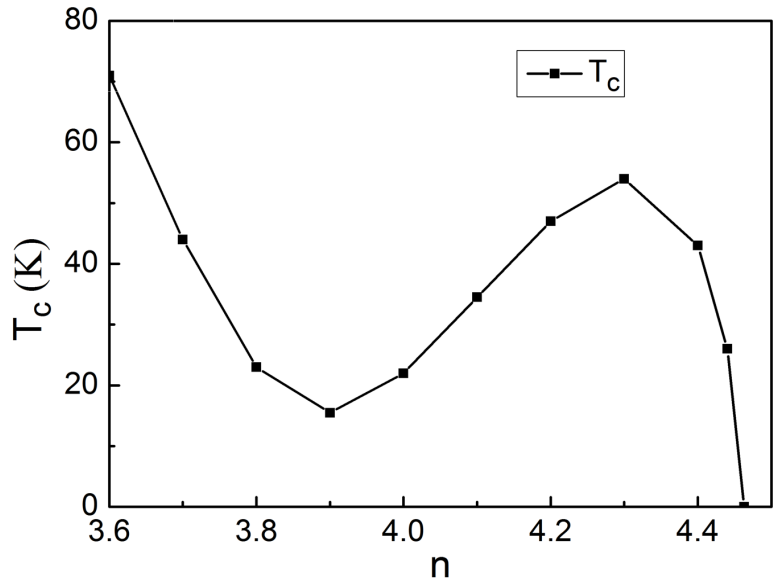

**Fig. 2.** The electron concentration *vs* temperature phase diagram of our model. The curve is the SC critical temperature as a function of electron occupation *n*.

The SC pairing symmetry and amplitude are dominated by the electron states near the FS. To further uncover the microscopic origin of the two-dome SC structure, we calculate the density of states (DOS) of the FS in the normal state, as shown in Fig. 3. In the entire SC region of $3.6 \leq n \leq 4.46$, the *xz* and *yz* orbitals contribute large DOS. The DOS sharply falls when $n > 4.46$, simultaneously, the SC order vanishes, as seen in Fig. 3. This suggests that the occurrence of the SC pairing is mainly associated with the variation of the weights of the *xz/yz* orbitals. This is also closely related to the disappearance of zone-center hole FS, as shown later in Fig.4.

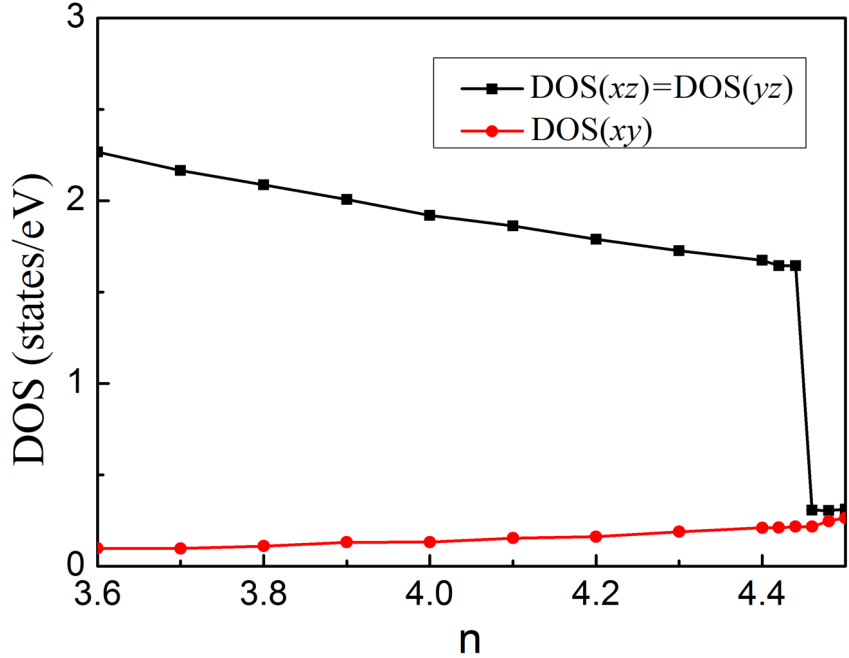

**Fig. 3.** The orbital-resolved DOS at the FS of the normal state as functions of electron occupation *n*.

Then we make a concrete analysis on the orbital weights and the orbital-resolved FS in the normal state. As shown in Fig. 4, the FS at $k_z = 0$ are contributed by the electronic states in *xz*, *yz* and *xy* orbitals. According to the colours of the curves representing the main orbital constitute, one finds that the two hole pockets in the center of Brillouin zone come from the mixture of *xz* and *yz* orbitals; the left/right zone-edge electron pocket come from the dominant *xz* and a little *xy* orbitals, and the up and down ones from the dominant *yz* and little *xy* orbitals. Thus in the first SC dome in the filling range of $3.6 \leq n \leq 3.9$, where is contributed mainly by the two circles in the zone center, the inter-orbital pairing is dominated by the *xz* and *yz* orbitals. As *n* increases, the sizes of the two hole FS circles in the zone center shrink. Accordingly, the hole number participating in the inter-orbital pairing decreases, so the critical temperature $T_c$ also decreases, as seen in Fig. 2. Similarly, another SC dome in the filling range of $3.9 \leq n \leq 4.46$, where the intra-orbital pairing is dominated by the *xz* and *yz* orbitals, is mainly contributed by the two electron FS on the zone edge.

When the electron occupancy varies from $n = 3.9$ to $n = 4.3$, the size of the electron FS continuously enlarges, which is favorable for the intra-orbital pairing, hence $T_c$ monotonically lifts. When $n$ further increases, the ratio of the $xy$ orbital on electron FS increases, and the two hole FS in the zone center disappear gradually, which is unfavorable for the inter-orbital pairing. Thus the intra-orbital SC pairing reaches the maximum at $n = 4.3$, as shown in Fig. 2. As $n$ increases from 3.6 to 4.46, the SC pairing changes from inter-orbital to intra-orbital. In this variation, the mixing $xz$ and $yz$ inter-orbitals pairing displays high symmetry, while the intra-orbital pairings of the $xz$ or $yz$ orbitals shows low symmetry, therefore, the rotational symmetry of SC pairing is lowered. This is consistent with the variation of the FS: in the former there exist two kind FS, electron-type and hole-type; with the increase of the doped electron number, the central hole FS is gradually filled, continuously shrinks until vanishes, while the electron FS continuously enlarges.

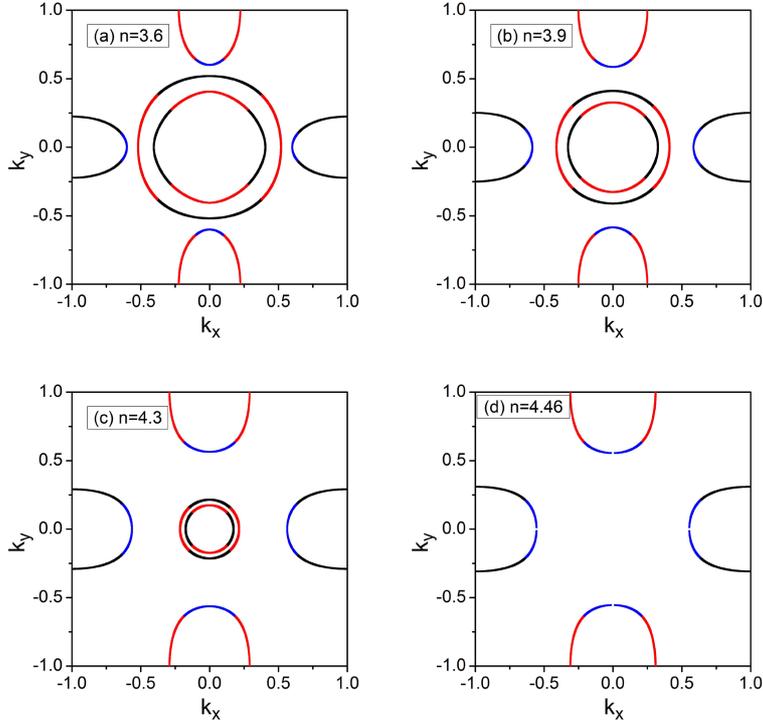

**Fig. 4.** The evolution of FS in the normal state at (a) $n$=3.6, (b) $n$=3.9, (c) $n$=4.3 and (d) $n$=4.46. The black, red and blue lines represent the FS dominantly contributed by $xz$, $yz$ and $xy$ orbitals, respectively.

We also present the dependence of the order parameter $V_c$ on the electron occupation $n$ in Fig. 5. Though the order parameter $V_c$ is defined as a $3 \times 3$ matrix in Eq. (4), our numerical calculations show that only the diagonal elements are finite, and these diagonal elements of $V_c$ linearly decline with the increasing electron density $n$. In the present theoretical framework, $V_c$ modulates the single electron potential of the charge density. This indicates that the order parameter $V_c$ plays the role of renormalized energy level, which can be absorbed into the single-electron energy levels.

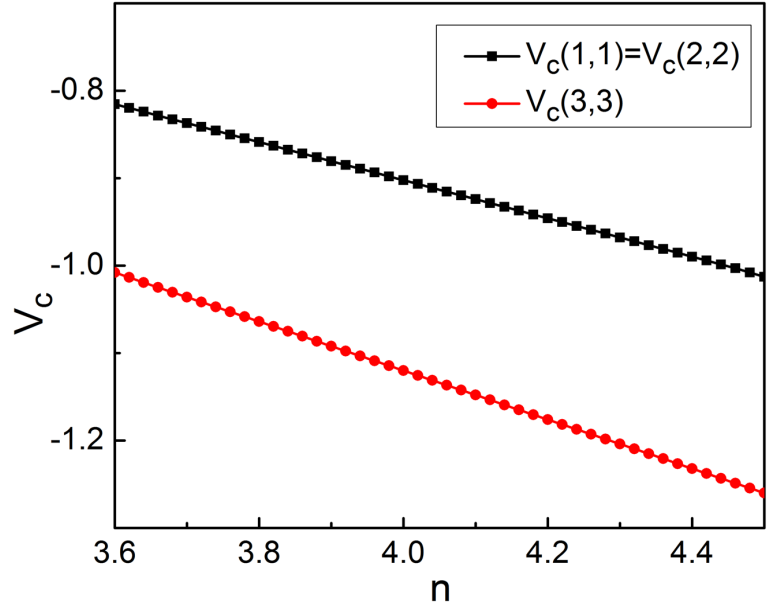

**Fig. 5.** The order parameters $V_c$ as functions of electron occupation $n$ at the temperature 5K.

## 4. Discussions

In the preceding study we have discussed the specific exchange interactions with the diagonal elements less than the off-diagonal ones. A natural question is what happens if the magnitudes of the diagonal and off-diagonal matrix elements of $J_1$ and $J_2$ are inverted. To clarify it, we also study another two typical cases with (a) $J_1 = \begin{bmatrix} 0.03 & 0.03 & 0.03 \\ 0.03 & 0.03 & 0.03 \\ 0.03 & 0.03 & 0.03 \end{bmatrix}$ eV, $J_2 = \begin{bmatrix} 0.075 & 0.075 & 0.075 \\ 0.075 & 0.075 & 0.075 \\ 0.075 & 0.075 & 0.075 \end{bmatrix}$ eV; and (b) $J_1 = \begin{bmatrix} 0.04 & 0.02 & 0.02 \\ 0.02 & 0.04 & 0.02 \\ 0.02 & 0.02 & 0.02 \end{bmatrix}$ eV, $J_2 = \begin{bmatrix} 0.1 & 0.05 & 0.05 \\ 0.05 & 0.1 & 0.05 \\ 0.05 & 0.05 & 0.05 \end{bmatrix}$ eV. The $n$-dependences of the order parameters are shown in Fig. 6(a) and Fig. 6(b), respectively. In Fig. 6(a), when the inter-orbital pairings strengths are equal to the intra-orbital ones, a similar two-dome SC structure with that in Fig. 1 emerges. The enhancement of the intra-orbital pairings in $xz$ and $yz$ orbitals increases the corresponding SC orders, so the boundary between the two SC domes shifts to low doping, and the inter-orbital pairings are suppressed by the enhanced intra-orbital ones, as shown in Fig. 6(a). When the intra-orbital spin coupling further enhances to the case (b), the SC phase with intra-orbital $s_{x^2y^2}$ and $s_{x^2+y^2}$ pairing further expands, as shown in Fig. 6(b). It seems that the two-dome SC structure is favored by the off-diagonal matrix elements of the spin coupling $J_1$ and $J_2$ being larger than the diagonal ones. In addition, the situation that the magnitude of $J_2^{\alpha\beta}$ is comparable with $J_1^{\alpha\beta}$ and $t_1$ can be found in the IBSC parent phases $SrFe_2As_2$ and $FeTe$ [60]. However, the parent compounds with $J_2^{\alpha\beta} > J_1^{\alpha\beta}$ are not definitely found yet, since it is hard to determine the multiple components of $J_2^{\alpha\beta}$ and $J_1^{\alpha\beta}$ in antiferromagnetic metals.

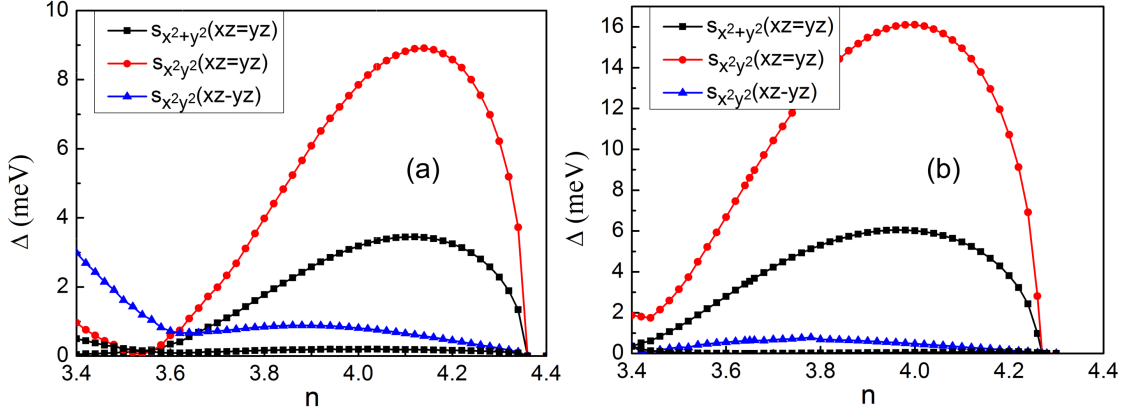

**Fig. 6.** The SC order parameters as functions of electron occupation $n$ at the temperature 5K. The color lines represent the same SC pairings as that in Fig. 1.

Similarly, adopting the spin-fluctuation exchange as the SC driving mechanism, the evolution of pairing symmetry and gap structure with doping has also been studied in earlier papers by S. Maiti *et al.* [14]. They showed that from small to moderate doping, the SC pairing favors a $s^{+-}$ gap, and a competition between the $s^{+-}$ symmetric and $d_{x^2-y^2}$ symmetric SC gaps arises in heavy electron doping. Though the doping evolution of the FS is consistent with Maiti *et al.*'s, we find that in our self-consistent theory the $d_{x^2-y^2}$-wave component is rather small, we focus our attention on the $s^{+-}$ and $s^{++}$ symmetric in the SC gap functions. We also find that the intra-orbital pairing favors a nodeless $s^{+-}$ wave gap function , which is consistent with the results in Ref. [10].

Also, as seen in Eq. (6), the $s^{++}$ and $d_{x^2-y^2}$ waves are associated with the NN vectors $\mathbf{e}_x$ and $\mathbf{e}_y$, while the $s^{+-}$ wave is associated with the NNN vector $\mathbf{e}_x + \mathbf{e}_y$. Hence, the $s^{++}$ and $d_{x^2-y^2}$ wave pairings are mainly determined by the parameter $J_1$, while the $s^{+-}$ wave pairing mainly depends on the parameter $J_2$. When $J_1$ (or $J_2$) becomes large or small, the $s^{++}$ and $d_{x^2-y^2}$ wave (or the $s^{+-}$ wave) pairings become strong or weak, respectively.

Meanwhile, we notice that although the present three-orbital TB model gives the correct FS topology of LaOFeAs [10], the hopping parameters are different from the five-orbital model [17-19], which could bring some numerical deviation. For example, the central hole FS pocket in Fig. 4(a) is considerably larger than observed in the ARPES experiment. There are also some other three-orbital TB model for iron-based superconductors [61-65], part of which are summarized in reference [66]. We also note that the present exchange parameters $J_1$ and $J_2$ adopted are slightly larger than those compounds with present three-orbital TB parameters, the present three-orbital SC theory is reasonable and meaningful so long as $J_1$ and $J_2$ are in the physical parameter range of iron-based SC. In addition, we do not consider the competition or coexistence of the SC and antiferromagnet orders. Meanwhile, without the single-occupancy constraint on the hopping term of the 'truncated' $t - J_1 - J_2$ model [15], the kinetic energy in Eq.(1) becomes of for free electrons and double occupancy is allowed. More charge transfer channels could exist, equivalent to the increase of the matrix elements of $T^{\alpha\beta}(\mathbf{k})$, hence, more large $J_1$ and $J_2$ are needed to stabilize the SC pairing state. These defects being improved in future study may change the phase boundary quantitatively, however, it does not qualitatively alter our results.

On the other hand, in the realistic experimental case, applying pressure leads to a series of complicated changes in material, such as the crystal structure, electron concentration, the strength of interactions, and so on. At present, we have shown that the variation of electron occupation $n$ leads to the two-dome SC phase diagram when the multiorbital material has proper exchange interaction to contribute the competing pairing forces. Therefore, one expects that the applying pressure may have the similar effect to doping, driving the appearance of two-dome SC phases in multiorbital superconductors.

## 5. Conclusions

In summary, we have shown that in the three-orbital TB model with appropriate values of the exchange coefficients $J_1^{\alpha\beta}$ and $J_2^{\alpha\beta}$, the orbital variation by doping drives a two-dome superconducting phase diagram. The two-dome structure mainly comes from the coexistence and competition between the inter-orbital and intra-orbital pairing of $xz$ and $yz$ orbitals. Such a competition widely exists in in multiorbital unconventional superconductors, therefore we believe that the two-dome superconducting phase diagram may be a common phenomenon.

## Acknowledgement

This work is supported by the National Natural Science Foundation of China under Grant Nos. 11774350, 11474287, and 11534010. Numerical calculations were performed at the Center for Computational Science of CASHIPS.